\documentclass[aps,pre,twocolumn,preprintnumbers,amsmath,amssymb]{revtex4}
\input epsf \usepackage[pdftex]{graphicx}
\usepackage{dcolumn}
\usepackage{bm}

\begin{document}

\title{Nucleation of cracks in a brittle sheet}   

\author{Cristiano L. Dias$^1$}\email{diasc@physics.mcgill.ca}
\author{ Jens Kr\"oger$^2$}
\author{ Daniel Vernon$^2$}
\author{ Martin Grant$^2$}  
\affiliation{ $^1$Department of Applied Mathematics, The University of
Western Ontario, Middlesex College, 1151 Richmond St. N., London (ON),
Canada N6A 5B7 \\
$^2$Physics Department, Rutherford
Building, McGill University, 3600 rue University, Montr\a'eal,
Qu\a'ebec,  H3A 2T8 Canada}

\date{\today}

\begin{abstract}
We use molecular dynamics to study the nucleation of cracks in a two
dimensional material without pre-existing cracks. We study models with zero and
non-zero shear  modulus. In both situations the time required for crack
formation obeys  an Arrhenius law, from which the energy barrier and
pre-factor  are extracted for different system sizes. For large
systems, the  characteristic time of rupture is found to decrease with
system  size, in agreement with classical Weibull theory. In the case
of zero shear modulus, the energy opposing rupture is identified with the
breakage of a single atomic layer. In the 
case of non-zero shear modulus, thermally activated fracture can only  be
studied within a reasonable time at very high strains.  In this
case the energy barrier involves the stretching of bonds within
several layers, accounting for a much higher barrier compared to the
zero shear modulus case. This barrier is understood within adiabatic
simulations.
\end{abstract}

\maketitle

\section{Introduction}

While our current understanding of fracture begins with the ideas of 
Griffith in 1921 \cite{GRIF21},  the study of its atomic mechanism has
attracted a large amount of attention in recent  years.  For example,
corrections to Griffith's results for a crack in a brittle material have been
proposed and verified with atomistic simulations
\cite{IPPO06,MATT05,MARD04}. Also,  large scale simulations  have been used
to study dynamical fracture \cite{BUEH03,HOLL98,BUEH07}. This
increase in interest in fracture is partly due to computer simulations
which promise  an understanding at the atomic level description of the
phenomena. However simulations face a fundamental problem
\cite{ALAV06,GOLU91}: many atomic deformations are thermally
activated and therefore involve long timescales which are difficult to
simulate.

Most simulations overcome this problem by studying fracture with a
pre-existing crack. In that  case, crack growth is a driven phenomena and
there is no energy barrier to be overcome.  Only a few simulations
have been used to study the formation of  cracks at non-zero finite
temperature without pre-existing cracks: void  formation
has been observed in 3D simulations of strained binary Lennard-Jones
systems \cite{LORE03}, and simulations for the rate of crack
nucleation have been performed in a 2D spring network \cite{SANT03}.
Experimentally, the rate of crack nucleation in heterogeneous
materials has been found to obey an Arrhenius law  with an energy
barrier scaling according to Griffith's results
\cite{GUAR99,RABI04}.  

In the present work,  we address the nucleation
of cracks in a brittle two-dimensional material, i.e.\ a sheet with a
thickness of one atomic layer,  through Langevin dynamics. The 
rate constant for the nucleation of cracks follows
an Arrhenius law, from which the energy barrier is extracted. Two
situations in a square lattice are studied:~atomic interactions
restricted to first neighbours and interactions extended to second
neighbours. In the former case, the shear modulus of the solid is zero
and the energy barrier is shown to be independent of system size:~the
breakage of a single bond propagates to the rest of the solid without
any cost. In the latter case the shear modulus is non-zero and a
finite size crack has to nucleate before rupture can propagate throughout 
the system. These two situations will be referred to as chain-like and 
solid-like models, respectively. 

This paper is organized as follows. In the next section we describe the 
models used in this paper followed by a description of how simulations
are carried out. In section \ref{section:results} we present the results for
the chain-like and solid-like models. The latter is physically more 
relevant to describe brittle materials and we discuss its energy barrier
in the context of Griffith theory in section \ref{section:griffith}. Our 
conclusions finalize the paper.

\section{Model}

\begin{figure}
\centerline{\includegraphics[height=40mm]{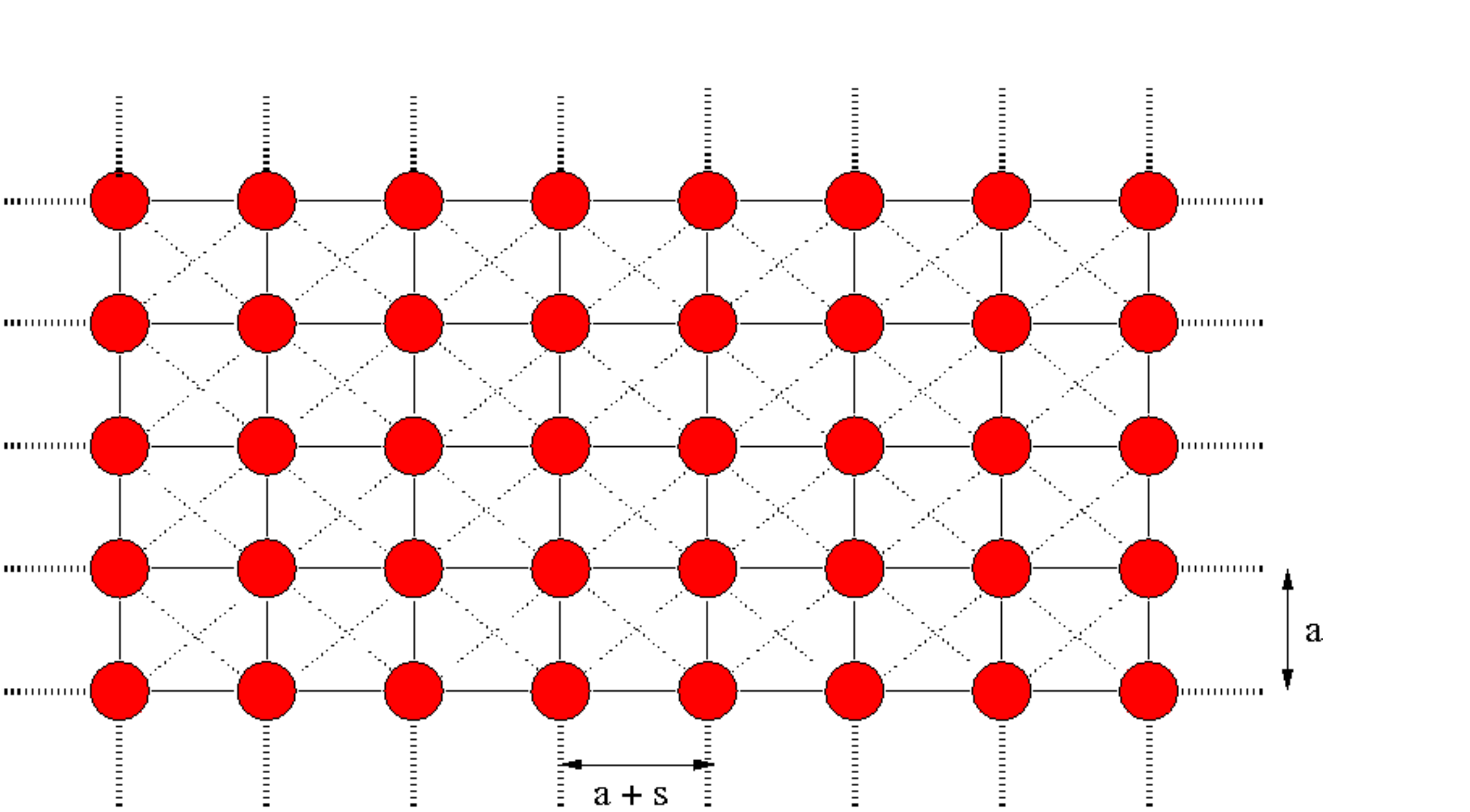} }
\caption{(Color Online) Schematic representation of system containing $M \times N = 5
\times 8$ atoms.  Atoms are only allowed to move in the horizontal
direction. First-neighbor interactions  are represented by straight
lines while dotted lines represent second-neighbor
interactions. Periodic boundary conditions are represented by dashed
lines.}
\label{fig:model}
\end{figure}

A stretched one dimensional chain has been previously used as a 
simple model for breakage of 
polymers~\cite{OLIV94,OLIV98,DIAS05,SAIN06,SAIN04,PUTH02}. Here 
we extend 
this model to study fracture in 2D brittle solids by bonding
the chains to each other such as to form a square lattice 
-- see Fig. \ref{fig:model}. We study samples containing 
$M$ chains which are made of $N=100$ atoms each. Those chains 
are stretched in the horizontal direction:~their constant 
length is $N(a+s)$, where $a$ is the equilibrium bond length 
and $s$ is the applied strain. By constraining the dynamics 
of atoms along the applied strain the system cannot form 
topological defects and the only mechanism for stress relaxation 
is fracture. Also, by choosing the constraint along the applied 
strain, we expect to be sampling the meaningful pathway for 
fracture while speeding up the simulation time. This is verified 
later in section~\ref{section:results} where we 
perform one set of simulations without this constraint.

A square lattice can be made isotropic by choosing the elastic constant 
between first neighbors to be twice as large as the elastic constant 
between second neighbors~\cite{MONE94}. To fulfil this condition we chose the 
following form for the Lennard-Jones potential between first neighbors:
\begin{equation}
V_f(r)=\epsilon[(a/r)^{12} -2(a/r)^6],
\label{eqn:lennardjones}
\end{equation} 
and
\begin{equation}
V_s(r) = 4 \epsilon_{s} [(a_{s}/r)^{12} - (a_s/r)^6],
\end{equation}
for second neighbors. We use $\epsilon = 1$ and $\epsilon_s=36/228
\epsilon$ for the binding energies and  $a = 1$ and $a_s = \sqrt{2} a$
for the equilibrium lengths.

The dynamics  of this system are obtained by solving a set of Langevin
equations for the position $x_{i,j}$ of each atom:
\begin{eqnarray}
m\frac{d^2 x_{i,j}}{dt^2}=& \sum_{k,l} F(x_{i,j}-x_{k,l}) - \eta
\frac{d x_{i,j}}{d t}  + f_{i,j}(t)
\label{eqn:Langevin}
\end{eqnarray}
where $F(x)$ is the force computed from the potential, $m$ is the
atomic mass and $\eta$ is the friction coefficient. The random force
$f_{i,j}(t)$ is related to $\eta$ by the fluctuation-dissipation
theorem.

Periodic boundary conditions in the horizontal direction imply
$x_{0,j}=x_{N,j}$ and $x_{N+1,j}=x_{1,j}$ for all $j$. Periodicity is
also imposed in the vertical direction to ensure that all chains are
equivalent: $x_{i,0}=x_{i,M}$  and $x_{i,M+1}=x_{i,1}$.  For
simplicity we use reduced units. Energy is given in units of
$\epsilon$, distance is given in terms of $a$, and time is given in
units of the smallest phonon oscillation period $P = {2
\pi}/({12\sqrt(2 \epsilon / m a^2)})$ of an intact chain
\cite{SAIN06}. Mass is written in terms of $m$ and the friction
coefficient is tuned to $\eta = 0.25(2\pi / P)$.

Initially all the horizontal bonds have the same length $a+s$ and all
vertical bonds are at their equilibrium length $a$. The velocity of
each atom is chosen randomly according the Boltzmann distribution. The
dynamics of the system are obtained by solving numerically the set of
equations \ref{eqn:Langevin} using the velocity-verlet algorithm
\cite{ALLE90} until the solid ruptures.

\section{Simulation}

\begin{figure}
\vspace{-0.5cm}
\centerline{\includegraphics[height=75mm]{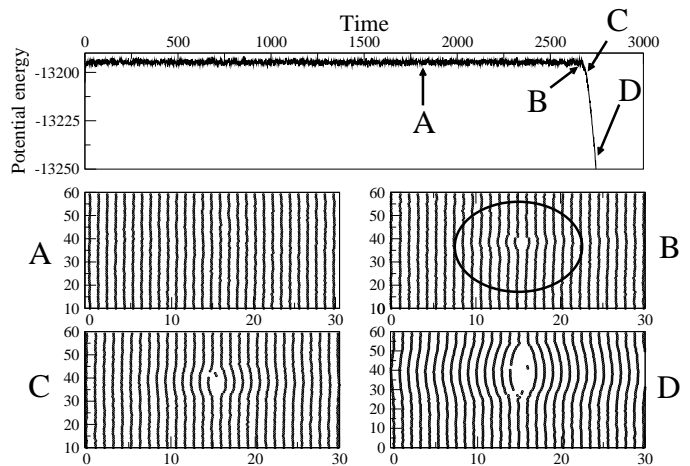} }
\caption{(UPPER PANEL) Dependence of the potential energy for the solid-like model defined by  $M \times N = 60 \times 100$, $S=0.065$ and $T=0.016$. Arrows indicate the instances at which the atomic configurations in panels A,B,C and D are shown.}
\label{fig:config}
\end{figure}

In Fig.~\ref{fig:config} (upper panel) we show the time dependence of
the potential energy for the solid-like model. This energy fluctuates
around its initial value showing that the initial stretched state is
a configuration at a local energy minimum. Rupture occurs at about $2750$ units of
time when the energy of the system drops abruptly. This shows that the
system is been driven towards an equilibrium state with lower
energy. In panels A,B,C and D of this figure, we shown atomic
configurations at different instances along fracture. In those panels,
the incipient crack is seen to propagate perpendicularly to the direction of
applied strain. This indicates that we can use the sum of bond 
lengths along the pathway where fracture is taking place as an order parameter 
$\phi$ for fracture. For convenience, 
the sum is taken over all the largest bonds percolating vertically along the 
sample and only the horizontal bonds are considered in the sum. Thus, 
initially $\phi = M(a+s)$ and $\phi$ increases until two surfaces are formed.

Notice that the potential energy in the upper panel of
Fig.~\ref{fig:config} shows no apparent precursor behavior for
rupture. Also, the energy barrier that the system has to overcome for
rupture to proceed is smaller than fluctuations in the total potential
energy. Thus it cannot be easily extracted  from an analysis of the
potential energy. To obtain this barrier we study the kinetics
of the system as it proceeds towards rupture. In particular we measure
the characteristic time of rupture and analyze this quantity from the
point of view of thermally activated systems.

To compute the characteristic time of rupture $\tau$ we use an
ensemble containing $S_0 = 1000$ samples for the chain-like model and
$S_0=500$ samples for the solid-like model. Those samples differ from
each other by the sets of initial velocities and random forces
$f_{i,j}(t)$. We will need to choose a value of the order parameter
$\phi$ which we will associate with irreversible rupture.  
The characteristic time for the incipient crack to
reach a particular size is computed by tracking the number of chains
$S (\phi, t)$ whose order parameter has not yet reached the value
$\phi$  at time $t$. For a fixed value of $\phi$ this function 
decreases exponentially with time, $S(\phi, t)=S_0 \exp (-t/\tau(\phi))$. 
The characteristic time  $\tau(\phi)$ depends on $\phi$ and  
is obtained from a fit of $S(\phi,t)$ to the numerical data.

\begin{figure}
\centerline{\includegraphics[height=70mm]{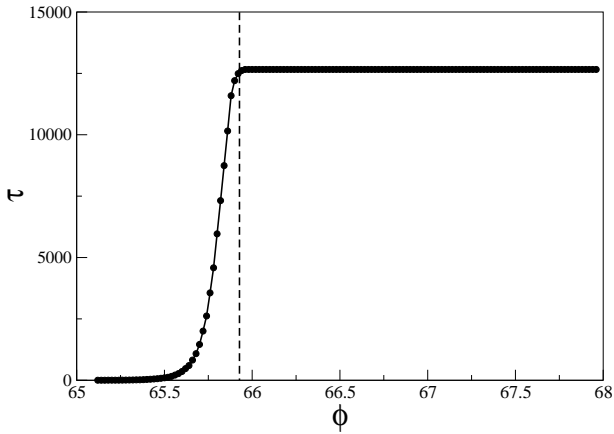} }
\caption{Dependence of the characteristic time on the cut-off value of
the order parameter for the solid-like model defined by $M \times N =
60 \times 100$, $S=0.065$ and $T=0.016$. The dashed line  separates
reversible (left side) from irreversible (right side) rupture.}
\label{fig:quebra}
\end{figure}

The time that characterizes rupture in an irreversible manner depends on 
the arbitrary choice of the cutoff value $\phi_c$ which we associate 
with rupture. To chose this cut-off we show in 
Fig.~\ref{fig:quebra} the dependence of $\tau$ on $\phi$. Two distinct 
regimes are apparent. The first regime occurs
when $\phi$ is smaller than ~$\sim 65.8$ (in units of $a$). In this regime,
$\phi$ increases very slowly with time.  The
underlying physics of this regime is the competition between thermal
fluctuations, which are responsible for increasing crack length, and
the restoring force on the atomic bonds. The second regime occurs when
the order parameter $\phi$ is greater than 65.8. Here, $\tau(\phi)$ has
reached a plateau, and $\phi$ increases very rapidly with time. Stress
relief of the material's bulk is the driving  force of this regime
which requires a larger crack and therefore produces a fast increase
in $\phi$:~irreversible rupture has occurred. So, from
Fig.~\ref{fig:quebra} we can determine the value of $\phi$ for which
rupture becomes irreversible. This value is $\phi_c=65.8$.

\section{Results \label{section:results}}

The nucleation of cracks can be thermally activated
\cite{GOLU91,ALAV06,SANT03} such that their occurrence is typical of an
Arrhenius process. Mathematically the characteristic time of rupture
$\tau$, the inverse of the nucleation rate, reads:
\begin{equation}
\tau=\tau_0\exp\Big( E_b / k_b T \Big)
\label{eqn:arrhenius}
\end{equation}
where $k_b T$ is thermal energy, $E_b$ is the energy barrier the
system has to overcome, and $\tau_0$ is the inverse of the attempt
frequency to rupture. The attempt frequency depends on the vibration
frequency of the system in the metastable wells of the energy
landscape through the local curvature of the energy
surface\cite{DALD99,ANGE00}. It also depends on the
friction coefficient in the Langevin equation \cite{KRAM40}. 
In this section we study the dependence of
$\tau$ on temperature for the chain-like and solid-like models to
extract both $E_b$ and $\tau_0$, which are intrinsic quantities of the
system being studied.

\subsection{Chain-like model}

\begin{figure}
\centerline{\includegraphics[height=70mm]{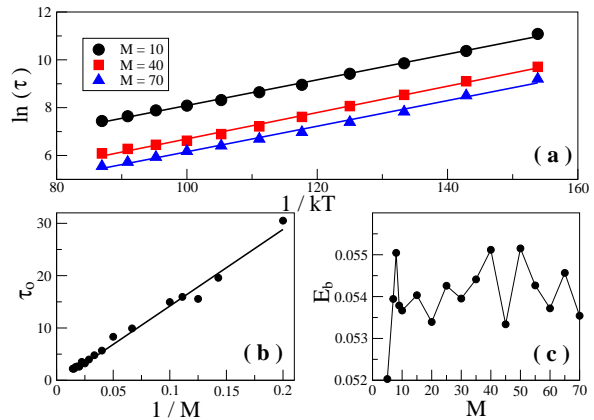} }
\caption{(Color online) CHAIN-LIKE MODEL -- (a) Dependence of $ln (\tau)$ on the
inverse of temperature for different system sizes $M$.
(b) Dependence of $\tau_0$ on system size. (c) Dependence of the
energy barrier $E_b$ on system size.}
\label{fig:scaling}
\end{figure}

In Fig.~\ref{fig:scaling} we show the temperature dependence of $\tau$
for different system sizes $M$. The strain of this system is set to
$s= 0.05$. For each system size, $\tau$ increases exponentially with
$1/k_bT$ -- in agreement with Eqn.~\ref{eqn:arrhenius}. The energy
barrier $E_b$ and the pre-factor $\tau_0$ were extracted from fits to
those results. Changing system size strongly affects the pre-factor
but has no effect, within error bars, on the energy barrier -- as can
be seen in Figures \ref{fig:scaling}(b) and (c) respectively.

The dependence of the pre-factor on system-size can be understood
qualitatively within the scope of a nucleation theory for
fracture. Intuitively, $\tau_0$ is proportional to the average time
between two consecutive attempts to nucleate a crack. Assuming that
cracks can nucleate in each of the $M$ chains of the system, then
$\tau_0 = 1 / M $ \citep{BAZA98,WEIB39}. Fig.~\ref{fig:scaling}(b)
shows the good agreement of simulation with this inverse relation.

The energy barrier can be understood quantitatively by assuming that
parallel chains are independent from each other. Under this
assumption, the energetic cost $E$ of elongating one atomic bond in a
single layer is given by \cite{OLIV94,DIAS05,SAIN06}:
\begin{equation}
E(\phi)=V(a+s+\phi) + (N-1) V \Big( a+s-\frac{\phi}{(N-1)} \Big)
\label{eqn:barrier_oliveira}
\end{equation}
where $\phi$ is the deviation of the broken bond length from its
strained elongation and $V(x)$ is the potential energy of an atomic
bond. Equation \ref{eqn:barrier_oliveira} corresponds to the sum of
potential energy of all the bonds in the layer precursor of
fracture. This equation considers that while one of the bonds
increases towards rupture by an amount $\phi$, the other bonds of the
same layer relax by an amount $\phi / (N-1)$ towards their
equilibrium value. For the parameters used in Fig.~\ref{fig:scaling},
i.e.\ $N=100$ and $s=0.05$, Equation \ref{eqn:barrier_oliveira}
predicts an energy barrier of $0.0564$. This is in good agreement with
our simulations where the barrier is approximately $0.054$ for all
system sizes.

Independence of parallel chains is the key assumption to explain
fracture in the chain-like model. This assumption can be understood as
follow. Since the shear modulus of this model is zero, no energetic
cost is associated with sheared configurations in the linear
regime. Therefore, an individual atomic layer can proceed towards
fracture independently of neighbouring layers until non-linear effects
become relevant. The energy barrier opposing this process is related
to the cost of increasing the length of one of the bonds of the layer --
independently  of other layers. Only when this bond becomes large
enough, neighbouring layers are driven towards rupture in a domino-like
process.

Eqn.~\ref{eqn:barrier_oliveira} results from the competition between
the energetic cost of extending one bond length of the chain and the
energetic gain of relaxing the remaining bonds. This contrasts with
Griffith's calculation where the barrier is related to the cost of
creating more surface and the energetic gain due to reducing the
strain in the bulk of the
material. Thus, despite its use in the literature, the square lattice
with only first neighbour interactions is a poor model for
fracture in a solid and Griffith's theory does not apply to this system.

\subsection{Solid-like model}

\begin{figure}[t]
\centerline{\includegraphics[height=70mm]{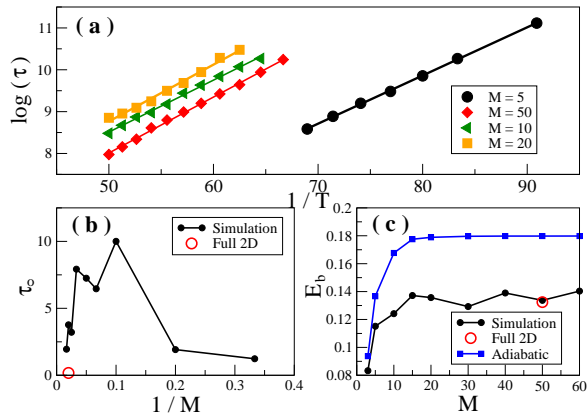} }
\caption{(Color online) SOLID-LIKE MODEL. (a) Dependence of $ln (\tau)$ on the
inverse of temperature for different system sizes $M$. (b) Dependence
of $\tau_0$ on system size.  (c) Dependence of the energy barrier
$E_b$ on system size. The results for the ``full 2D'' system correspond 
to a dynamically unconstrained model -- see text for more details.}
\label{fig:scaling2}
\end{figure}

We now study the solid-like model. In Fig.~\ref{fig:scaling2} we show
the temperature dependence of $\tau$  for different system sizes $M$
and applied strain $s= 0.065$. As in the previous model we study this
system by fitting the time of fracture to Eqn.~\ref{eqn:arrhenius},
obtaining the energy barrier $E_b$ and the pre-factor $\tau_0$ for
each system size. Those results are shown in
Fig.~\ref{fig:scaling2}(b-c).

The pre-factor, Fig.~\ref{fig:scaling2}(b), presents two  regimes: for
systems containing less than 15 chains, i.e.\ $M<15$, $\tau_0$
increases with system size; however for $M>15$, the pre-factor
decreases as system size increases. Those behaviors are related to finite
size effects. When $M<15$, the relaxation region around the crack 
is of the same size as the system. On
the other  hand, increasing the size of the solid above 15 layers implies
that more  nucleation sites are available for rupture, and $\tau_0$ decreases 
with $M$, as in the previous model \cite{BAZA98,WEIB39}. A quantitative 
assessment of the pre-factor would involve the
generalisation of Kramer's  result to higher dimensions \citep{LANG69}. 
This calculation was performed successfully to study rupture  
in a one-dimensional chain \cite{SAIN06} but its application to 
the present model  is beyond the scope of this paper.

In Fig.~\ref{fig:scaling2}(c) we show the energetic cost for
nucleating a crack in the solid-like model as a function of system
size $M$. For solids smaller than $M=15$, the energy barrier
increases considerably with system size: more than $150 \%$ in
Fig.~\ref{fig:scaling2}(c); while for solids larger than $M=15$, the
increase is only marginal and show a saturation trend at $E_b \sim 0.14$ -- 
indicating that finite-size effects become negligible. This value is comparable
to the adiabatic barrier. This barrier is computed by
extending the length  of one bond in small steps and
restraining it while the other bonds are relaxed at zero
temperature. In this process the energy increases until the critical
crack is formed. The maximum energy seen in
this process corresponds to the energy required to nucleate the crack
at zero temperature.  This adiabatic energy for the different system sizes are
represented by squares in Fig.~\ref{fig:scaling2}. Notice that the barrier 
obtained in our simulations is smaller than the 
adiabatic energy barrier by about 23 \%. A smaller simulated 
barrier compared to the adiabatic case has also been observed for 
one dimensional systems \cite{OLIV98}. A possible explanation for 
this discrepancy might be that a zero temperature calculation does 
not account for entropy which plays a role in the free energy opposing 
rupture in system with multiple degrees of freedom.

One important simplification imposed in our model with respect to
two-dimensional solids consists in constraining the dynamics of atoms
to one dimension. However by imposing this constraint along the direction of
applied stress, we  expect to be sampling the meaningful pathway for
rupture of a  2D-solid. To verify this statement we  performed a 
set of simulations on a $M \times N =50 \times 100$ system where the 
constraint on the motion of atoms was
removed. The results of those simulations are shown in
Fig.~\ref{fig:scaling2}(b-c) and are referred to as full 2D. Notice
that the full 2D system has a much lower pre-factor than our
constrained system. A discrepancy in the pre-factor is expected  since
it is related to the vibration of the system, and
therefore its dynamics, which is different in both models. However the
energy barrier of our constraint model and the full 2D are equal
within error bars. We are therefore confident  that our constrained
model can be used to study the energetic behavior of  2D solids. In
the next section we discuss the simulated energy barrier  in the
context of Griffith theory for rupture and adiabatic simulations.

\section{Solid-like model and Griffith \label{section:griffith}}

\begin{figure}[t]
\centerline{\includegraphics[height=70mm]{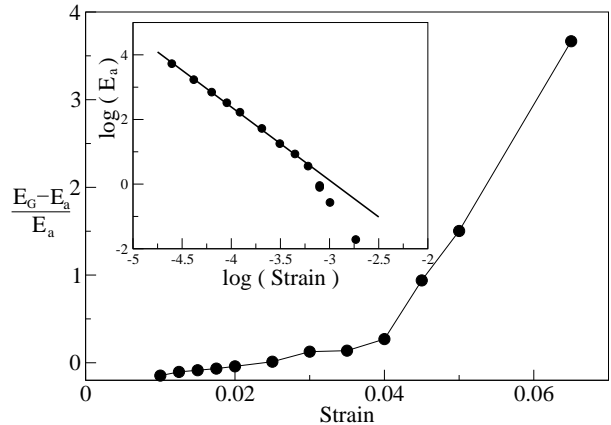} }
\caption{Relative difference in energy barrier as computed from
  Griffith calculation ($E_G$) and the adiabatic simulation
  ($E_a$):~$(E_G-E_a)/E_G$. INSET - dependence of the adiabatic
  barrier on strain in a log-log scale. The linear fit of systems with 
a strain smaller than 0.04 is shown.}
\label{fig:adiabatic}
\end{figure}

The introduction of a crack of size $L$ in a solid characterised by a
Young's modulus $E$ and subjected to a stress $\sigma$ will result
in a stress-energy relief of $\pi \sigma^2 L^2/2 E$. But this crack
will also involve a cost of $2\gamma L$, where $\gamma$ is a surface
energy, such that the dependence of the energy on the crack size $L$
is \cite{RABI04}:
\begin{equation}
E_G(L)= -\frac{ \pi L^2 \sigma^2}{2E} + 2 \gamma L.
\label{eqn:griffith}
\end{equation}
This potential energy reaches a maximum when $\partial E_G(L) /
\partial L = 0$. This occurs at the critical value $L_G=  2 E \gamma /
(\pi \sigma^2)$. Beyond this crack length, the crack propagates
spontaneously to reduce the bulk strain in the material until the solid is
broken in two  pieces. The barrier for crack nucleation occurs at this
critical  length:~$E_G(L_G)=\frac{2 \gamma^2 E}{\pi \sigma^2}$ or
$\frac{2 \gamma^2}{\pi s^2E}$. The Young modulus of the solid-like model 
is $E=77.91$  (in units of $\epsilon_f / a^3$) and the energy 
necessary to create two surfaces $2\gamma$ is equal to the energy 
of two weak and one strong bond per interatomic
distance:~$\gamma=0.6587$. For a strain $s=0.065$, the Griffith energy
barrier is $E_G=0.837$, that is, approximately six times the value
obtained from our statistical simulation. 

It is of no surprise that Griffith's calculation is not valid for
large strains. First, because in this highly stretched regime, linear
elasticity  theory is not valid. Second, for strains larger than 0.04  
only one atomic layer needs to be completely broken in order to 
initiate  the rupture process. In other words, while in
Griffith's regime the mechanism behind  rupture is the competition
between the formation of new surface and stress relaxation, the
physics of rupture in the highly stressed regime is the competition
between stress relaxation and bond stretching at the formation of the 
incipient crack. 

To understand the range of validity of Griffith calculation, we
performed adiabatic relaxation ($T=0$) where bonds were mathematically
cut within a line perpendicular to the applied strain. After cutting
those bonds, atoms were relaxed until the force on each one of them
was smaller than $1 \times 10^{-5}$. To avoid finite-size effects we
increased the size of the system from N=100 at S=0.04 to N=600 at
S=0.01. The relative energy barrier   computed from this process with
respect to Griffith barrier is shown  in
Fig. \ref{fig:adiabatic}. This Figure shows that for strains below
0.04, the adiabatic  barrier differs at most by 30 \% from the Griffith
barrier. On the other hand,  for strains beyond  0.04 those barriers
are several times different and this  difference increases with
strain. In the inset of Fig. \ref{fig:adiabatic} we also show in a
log-log scale the  dependence of the energy barrier on strain. For
strains below 0.04, those  quantities scale with an exponent  of
-2.26. This is very close to the exponent predicted by Griffith's
theory: -2. The behavior at higher strains deviate  from this scaling.
This clearly shows that for our model Griffith's theory is valid for
strains smaller than 0.04. We note that the calculation of the  energy
barrier with strain dependent Young modulus and surface tension --  as
described in Ref. \cite{MATT05} -- gave results in greater
disagreement than the ones of Griffith.

\section{Conclusion}

Due to its simple dynamics and large system size, our atomistic
simulation of  the nucleation of cracks in thin brittle sheets is an
ideal system for the  study of noise activated processes and
nucleation theory. In particular, we found  that the energy barrier
for crack nucleation in a square lattice with only  first-neighbor
interactions is comparable to the barrier of one-dimensional  chains
due to the zero shear modulus of this system. For the more interesting
case where second-neighbor interactions are incorporated into the
model, we found an agreement between the simulated energy barrier at 
high strains and the one computed from an adiabatic relaxation. This 
barrier involves several layers, accounting for a much 
higher barrier compared to the case of isolated chains. We believe 
that extensions of the present study such as to investigate 
nucleation of pre-existing cracks would be a valuable contribution to 
understand fracture.

{\bf Ackowledgements}

This work was supported by the Natural Sciences and Engineering Research Council of Canada, and \textit{le Fonds Qu\'eb\'ecois de la recherche sur la nature et les technologies}.



\begin{thebibliography}{21}
\expandafter\ifx\csname natexlab\endcsname\relax\def\natexlab#1{#1}\fi
\expandafter\ifx\csname bibnamefont\endcsname\relax
  \def\bibnamefont#1{#1}\fi
\expandafter\ifx\csname bibfnamefont\endcsname\relax
  \def\bibfnamefont#1{#1}\fi
\expandafter\ifx\csname citenamefont\endcsname\relax
  \def\citenamefont#1{#1}\fi
\expandafter\ifx\csname url\endcsname\relax
  \def\url#1{\texttt{#1}}\fi
\expandafter\ifx\csname urlprefix\endcsname\relax\def\urlprefix{URL }\fi
\providecommand{\bibinfo}[2]{#2}
\providecommand{\eprint}[2][]{\url{#2}}

\bibitem[{\citenamefont{Griffith}(1921)}]{GRIF21}
\bibinfo{author}{\bibfnamefont{A.~A.} \bibnamefont{Griffith}},
  \bibinfo{journal}{Philosophical Transactions of the Royal Society of London.
  Series A}
  \textbf{\bibinfo{volume}{221}}, \bibinfo{pages}{163} (\bibinfo{year}{1921}).

\bibitem[{\citenamefont{Ippolito et~al.}(2006)\citenamefont{Ippolito, Mattoni,
  and Colombo}}]{IPPO06}
\bibinfo{author}{\bibfnamefont{M.}~\bibnamefont{Ippolito}},
  \bibinfo{author}{\bibfnamefont{A.}~\bibnamefont{Mattoni}}, 
  \bibinfo{author}{\bibfnamefont{L.}~\bibnamefont{Colombo}},\bibnamefont{and}
\bibinfo{author}{\bibfnamefont{Nicola}~\bibnamefont{Pugno}},
  \bibinfo{journal}{Phys. Rev. B} \textbf{\bibinfo{volume}{73}},
  \bibinfo{pages}{104111} (\bibinfo{year}{2006}).

\bibitem[{\citenamefont{Mattoni et~al.}(2005)\citenamefont{Mattoni, Colombo,
  and Cleri}}]{MATT05}
\bibinfo{author}{\bibfnamefont{A.}~\bibnamefont{Mattoni}},
  \bibinfo{author}{\bibfnamefont{L.}~\bibnamefont{Colombo}}, \bibnamefont{and}
  \bibinfo{author}{\bibfnamefont{F.}~\bibnamefont{Cleri}},
  \bibinfo{journal}{Phys. Rev. Lett.} \textbf{\bibinfo{volume}{95}},
  \bibinfo{pages}{115501} (\bibinfo{year}{2005}).

\bibitem[{\citenamefont{Marder}(2004)}]{MARD04}
\bibinfo{author}{\bibfnamefont{M.} \bibnamefont{Marder}},
  \bibinfo{journal}{International Journal of Fracture} \textbf{\bibinfo{volume}{130}},
  \bibinfo{pages}{517} (\bibinfo{year}{2004}).

\bibitem[{\citenamefont{Buehler et al}(2007)\citenamefont{Buehler, Tang, Duin and Goddard III}}]{BUEH07}
\bibinfo{author}{\bibfnamefont{Markus~J.}~\bibnamefont{Buehler}},
  \bibinfo{author}{\bibfnamefont{Harvey}~\bibnamefont{Tang}}, 
\bibinfo{author}{\bibfnamefont{Adri~C.~T.}~\bibnamefont{van Duin}},
\bibnamefont{and}
  \bibinfo{author}{\bibfnamefont{William~A.}~\bibnamefont{Goddard III}},
  \bibinfo{journal}{Phys. Rev. Lett.} \textbf{\bibinfo{volume}{99}},
  \bibinfo{pages}{165502} (\bibinfo{year}{2007}).


\bibitem[{\citenamefont{Dominic Holland et al}(1998)}]{HOLL98}
\bibinfo{author}{\bibfnamefont{Dominic} \bibnamefont{Holland}}
  \bibnamefont{and} \bibinfo{author}{\bibfnamefont{M.}~\bibnamefont{Marder}},
  \bibinfo{journal}{Physical Review Letters} \textbf{\bibinfo{volume}{80}},
  \bibinfo{pages}{746} (\bibinfo{year}{1998}).



\bibitem[{\citenamefont{Markus, J.~Buehler, and Gao}(2003)}]{BUEH03}
\bibinfo{author}{\bibfnamefont{F.~F.~A.} \bibnamefont{Markus J.~Buehler}}
  \bibnamefont{and} \bibinfo{author}{\bibfnamefont{H.}~\bibnamefont{Gao}},
  \bibinfo{journal}{Nature} \textbf{\bibinfo{volume}{426}},
  \bibinfo{pages}{141} (\bibinfo{year}{2003}).



\bibitem[{\citenamefont{Zapperi}(2006)}]{ALAV06}
\bibinfo{author}{\bibfnamefont{Mikko~J.~Alava}},
 \bibinfo{author}{\bibfnamefont{Phani~K.~V.~V.~Nukala}},
 \bibnamefont{and}
\bibinfo{author}{\bibfnamefont{Stefano~Zapperi}}, 
\bibinfo{journal}{Advances in Physics}
  \textbf{\bibinfo{volume}{55}}, \bibinfo{pages}{349} (\bibinfo{year}{2006}).

\bibitem[{\citenamefont{Golubovic and Feng}(1991)}]{GOLU91}
\bibinfo{author}{\bibfnamefont{L.}~\bibnamefont{Golubovic}} \bibnamefont{and}
  \bibinfo{author}{\bibfnamefont{S.}~\bibnamefont{Feng}},
  \bibinfo{journal}{Physical Review A} \textbf{\bibinfo{volume}{43}},
  \bibinfo{pages}{5223} (\bibinfo{year}{1991}).


\bibitem[{\citenamefont{Lorenz and Stevens}(2003)}]{LORE03}
\bibinfo{author}{\bibfnamefont{C.~D.} \bibnamefont{Lorenz}} \bibnamefont{and}
  \bibinfo{author}{\bibfnamefont{M.~J.} \bibnamefont{Stevens}},
  \bibinfo{journal}{Phys. Rev. E} \textbf{\bibinfo{volume}{68}},
  \bibinfo{pages}{021802} (\bibinfo{year}{2003}).

\bibitem[{\citenamefont{Santucci et~al.}(2003)\citenamefont{Santucci, Vanel,
  Guarino, Scorretti, and Ciliberto}}]{SANT03}
\bibinfo{author}{\bibfnamefont{S.}~\bibnamefont{Santucci}},
  \bibinfo{author}{\bibfnamefont{L.}~\bibnamefont{Vanel}},
  \bibinfo{author}{\bibfnamefont{A.}~\bibnamefont{Guarino}},
  \bibinfo{author}{\bibfnamefont{R.}~\bibnamefont{Scorretti}},
  \bibnamefont{and}
  \bibinfo{author}{\bibfnamefont{S.}~\bibnamefont{Ciliberto}},
  \bibinfo{journal}{Europhys. Lett.} \textbf{\bibinfo{volume}{62}},
  \bibinfo{pages}{320} (\bibinfo{year}{2003}).


\bibitem[{\citenamefont{Guarino et~al.}(1999)\citenamefont{Guarino, Ciliberto,
  and Garcimart\a'in}}]{GUAR99}
\bibinfo{author}{\bibfnamefont{A.}~\bibnamefont{Guarino}},
  \bibinfo{author}{\bibfnamefont{S.}~\bibnamefont{Ciliberto}},
  \bibnamefont{and}
  \bibinfo{author}{\bibfnamefont{A.}~\bibnamefont{Garcimart\a'in}},
  \bibinfo{journal}{Europhys. Lett.} \textbf{\bibinfo{volume}{47}},
   \bibinfo{pages}{456} (\bibinfo{year}{1999}).


\bibitem[{\citenamefont{Rabinovitch et~al.}(2004)\citenamefont{Rabinovitch,
  Friedman, and Bahat}}]{RABI04}
\bibinfo{author}{\bibfnamefont{A.}~\bibnamefont{Rabinovitch}},
  \bibinfo{author}{\bibfnamefont{M.}~\bibnamefont{Friedman}}, \bibnamefont{and}
  \bibinfo{author}{\bibfnamefont{D.}~\bibnamefont{Bahat}},
  \bibinfo{journal}{Europhys. Lett.} \textbf{\bibinfo{volume}{67}},
  \bibinfo{pages}{969} (\bibinfo{year}{2004}).

\bibitem[{\citenamefont{Sain et~al.}(2006)\citenamefont{Sain, Dias, and
  Grant}}]{SAIN06}
\bibinfo{author}{\bibfnamefont{A.}~\bibnamefont{Sain}},
  \bibinfo{author}{\bibfnamefont{C.~L.} \bibnamefont{Dias}}, \bibnamefont{and}
  \bibinfo{author}{\bibfnamefont{M.}~\bibnamefont{Grant}},
  \bibinfo{journal}{Physical Review E} \textbf{\bibinfo{volume}{74}},
  \bibinfo{pages}{046111} (\bibinfo{year}{2006}).

\bibitem[{\citenamefont{Oliveira and Taylor}(1994)}]{OLIV94}
\bibinfo{author}{\bibfnamefont{F.~A.} \bibnamefont{Oliveira}} \bibnamefont{and}
  \bibinfo{author}{\bibfnamefont{P.~L.} \bibnamefont{Taylor}},
  \bibinfo{journal}{Journal of Chemical Physics}
  \textbf{\bibinfo{volume}{101}}, \bibinfo{pages}{10118}
  (\bibinfo{year}{1994}).

\bibitem[{\citenamefont{Oliveira}(1998)}]{OLIV98}
\bibinfo{author}{\bibfnamefont{F.~A.} \bibnamefont{Oliveira}},
  \bibinfo{journal}{Physical Review B}
  \textbf{\bibinfo{volume}{57}}, \bibinfo{pages}{10576}
  (\bibinfo{year}{1998}).

\bibitem[{\citenamefont{Dias et~al.}(2005)\citenamefont{Dias, Dube, Oliveira,
  and Grant}}]{DIAS05}
\bibinfo{author}{\bibfnamefont{C.~L.} \bibnamefont{Dias}},
  \bibinfo{author}{\bibfnamefont{M.}~\bibnamefont{Dube}},
  \bibinfo{author}{\bibfnamefont{F.A.}~\bibnamefont{Oliveira}}, \bibnamefont{and}
  \bibinfo{author}{\bibfnamefont{M.}~\bibnamefont{Grant}},
  \bibinfo{journal}{Physical Review E} \textbf{\bibinfo{volume}{72}},
  \bibinfo{pages}{011918} (\bibinfo{year}{2005}).



\bibitem[{\citenamefont{Sain and Wortis}(2004)}]{SAIN04}
\bibinfo{author}{\bibfnamefont{Anirban}~\bibnamefont{Sain}} \bibnamefont{and}
  \bibinfo{author}{\bibfnamefont{Michael} \bibnamefont{Wortis}},
  \bibinfo{journal}{Physical Review E}
  \textbf{\bibinfo{volume}{70}}, \bibinfo{pages}{031102} (\bibinfo{year}{2004}).



\bibitem[{\citenamefont{Rosabella and Sebastian}(2002)}]{PUTH02}
\bibinfo{author}{\bibfnamefont{Rosabella~K.}~\bibnamefont{Puthur}} \bibnamefont{and}
  \bibinfo{author}{\bibfnamefont{K.~L.~} \bibnamefont{Sebastian}},
  \bibinfo{journal}{Physical Review B}
  \textbf{\bibinfo{volume}{66}}, \bibinfo{pages}{024304} (\bibinfo{year}{2002}).


\bibitem[{\citenamefont{Monette and Anderson}(1994)}]{MONE94}
\bibinfo{author}{\bibfnamefont{L.}~\bibnamefont{Monette}} \bibnamefont{and}
  \bibinfo{author}{\bibfnamefont{M.~P.} \bibnamefont{Anderson}},
  \bibinfo{journal}{Modelling Simul. Mater. Sci. Eng.}
  \textbf{\bibinfo{volume}{2}}, \bibinfo{pages}{53} (\bibinfo{year}{1994}).

\bibitem[{\citenamefont{Allen and Tildesley}(1990)}]{ALLE90}
\bibinfo{author}{\bibfnamefont{M.~P.}~\bibnamefont{Allen}} \bibnamefont{and}
  \bibinfo{author}{\bibfnamefont{D.~J.} \bibnamefont{Tildesley}},
  \emph{Computer Simulation of Liquids}, (Clarendon, Oxford, 1990).
  

\bibitem[{\citenamefont{Kramer}(1940)}]{KRAM40}
\bibinfo{author}{\bibfnamefont{H.A.~}~\bibnamefont{Kramers}}
\bibinfo{journal}{Physics (Utrecht)}
\textbf{\bibinfo{volume}{7}}, \bibinfo{pages}{284} (\bibinfo{year}{1940}).

\bibitem[{\citenamefont{Angelani et~al.}(2000)\citenamefont{Angelani, Parisi,
  Ruocco, and Viliani}}]{ANGE00}
\bibinfo{author}{\bibfnamefont{L.}~\bibnamefont{Angelani}},
  \bibinfo{author}{\bibfnamefont{G.}~\bibnamefont{Parisi}},
  \bibinfo{author}{\bibfnamefont{G.}~\bibnamefont{Ruocco}}, \bibnamefont{and}
  \bibinfo{author}{\bibfnamefont{G.}~\bibnamefont{Viliani}},
  \bibinfo{journal}{Phys. Rev. E} \textbf{\bibinfo{volume}{61}},
  \bibinfo{pages}{1681} (\bibinfo{year}{2000}).

\bibitem[{\citenamefont{Daldoss et~al.}(1999)\citenamefont{Daldoss, Pilla,
  Viliani, and Brangian}}]{DALD99}
\bibinfo{author}{\bibfnamefont{G.}~\bibnamefont{Daldoss}},
  \bibinfo{author}{\bibfnamefont{O.}~\bibnamefont{Pilla}},
  \bibinfo{author}{\bibfnamefont{G.}~\bibnamefont{Viliani}}, 
  \bibinfo{author}{\bibfnamefont{C.}~\bibnamefont{Brangian}},\bibnamefont{and}
\bibinfo{author}{\bibfnamefont{G.}~\bibnamefont{Ruocco}},
  \bibinfo{journal}{Physical Review B} \textbf{\bibinfo{volume}{60}},
  \bibinfo{pages}{003200} (\bibinfo{year}{1999}).

\bibitem[{\citenamefont{Hanggi et~al.}(1990)\citenamefont{Hanggi, Talkner, and
  Borkovec}}]{HANN90}
\bibinfo{author}{\bibfnamefont{P.}~\bibnamefont{Hanggi}},
  \bibinfo{author}{\bibfnamefont{P.}~\bibnamefont{Talkner}}, \bibnamefont{and}
  \bibinfo{author}{\bibfnamefont{M.}~\bibnamefont{Borkovec}},
  \bibinfo{journal}{Reviews of Modern Physics} \textbf{\bibinfo{volume}{62}},
  \bibinfo{pages}{251} (\bibinfo{year}{1990}).

\bibitem[{\citenamefont{Bazant and Planas}(1998)}]{BAZA98}
\bibinfo{author}{\bibfnamefont{Z.~P.} \bibnamefont{Bazant}} \bibnamefont{and}
  \bibinfo{author}{\bibfnamefont{J.}~\bibnamefont{Planas}},
  \emph{\bibinfo{title}{Fracture and Size effect in Concrete and other brittle
  materials.}} (\bibinfo{publisher}{CRC Press. Boca Raton.},
  \bibinfo{year}{1998}).

\bibitem[{\citenamefont{Weibull}(1939)}]{WEIB39}
\bibinfo{author}{\bibfnamefont{W.}~\bibnamefont{Weibull}}, \bibinfo{journal}{Proc.
  Royal Swedish Academy of Eng. Sci.} \textbf{\bibinfo{volume}{151}},
  \bibinfo{pages}{1} (\bibinfo{year}{1939}).

\bibitem[{\citenamefont{Langer}(1969)}]{LANG69}
\bibinfo{author}{\bibfnamefont{J.}~\bibnamefont{Langer}},
  \bibinfo{journal}{Annals phys.} \textbf{\bibinfo{volume}{54}},
  \bibinfo{pages}{258} (\bibinfo{year}{1969}).


\bibitem[{\citenamefont{Pacia et~al.}(2005)\citenamefont{Pacia, Suna,
  Belytschkob, and Schatz}}]{PACI05}
\bibinfo{author}{\bibfnamefont{J.~T.} \bibnamefont{Pacia}},
  \bibinfo{author}{\bibfnamefont{L.}~\bibnamefont{Suna}},
  \bibinfo{author}{\bibfnamefont{T.}~\bibnamefont{Belytschkob}},
  \bibnamefont{and} \bibinfo{author}{\bibfnamefont{G.~C.}
  \bibnamefont{Schatz}}, \bibinfo{journal}{Chemical Physics Letters}
  \textbf{\bibinfo{volume}{14}}, \bibinfo{pages}{16} (\bibinfo{year}{2005}).

\end{thebibliography}

\end{document}